\documentclass[acmsmall,authorversion]{acmarttr}

\usepackage{pdfsync}
\usepackage{verbatim}
\usepackage{graphicx}
\usepackage{xspace}
\usepackage{multirow}
\usepackage{rotating}
\usepackage{amsmath}

\def\..{\,\mathpunct{\ldotp\ldotp}} 

\newcommand{\Z}{\mathbf Z}
\newcommand{\F}{\mathbf F}

\newcommand{\xorshift}[1][]{\texttt{xorshift#1}\xspace}

\newcommand{\xoroshiro}[1][]{\texttt{xoroshiro#1}\xspace}

\newcommand{\xoroshiropp}[1][]{\texttt{xoroshiro#1++}\xspace}

\newcommand{\xorshiftp}[1][]{\texttt{xorshift#1+}\xspace}

\newcommand{\xoshirop}[1][]{\texttt{xoshiro#1+}\xspace}
\newcommand{\xoshiropp}[1][]{\texttt{xoshiro#1++}\xspace}

\newcommand{\mt}[1][]{\texttt{MT19937}\xspace}

\begin{document}

\bibliographystyle{ACM-Reference-Format-Journals}

\title{It Is High Time We Let Go Of The Mersenne Twister}
\titlenote{This work has been supported by a Google Focused Research Award.}
\author{Sebastiano Vigna}
\affiliation{%
  \institution{Universit\`a degli Studi di Milano}
  \country{Italy}
}

\begin{abstract}
When the Mersenne Twister made his first appearance in 1997 it was a powerful example
of how linear maps on $\F_2$ could be used to generate pseudorandom numbers. In
particular, the easiness with which generators with long periods could be defined
gave the Mersenne Twister a large following, in spite of the fact that such long periods
are not a measure of quality, and they require a large amount of memory. Even
at the time of its publication, several defects of the Mersenne Twister were predictable,
but they were somewhat obscured by other interesting properties.
Today the Mersenne Twister is the default generator in C compilers,
the Python language, the Maple mathematical computation system, and in many other 
environments. Nonetheless, knowledge accumulated in the last $20$ years suggests that
the Mersenne Twister has, in fact, severe defects, and should never be used as a general-purpose 
pseudorandom number generator. Many of these results are folklore, or are scattered through very specialized
literature. This paper surveys these results for the non-specialist, providing new,
simple, understandable examples, and it is intended
as a guide for the final user, or for language implementors, so that they can take
an informed decision about whether to use the Mersenne Twister or not.
\end{abstract}

\maketitle





\section{Introduction}

``Mersenne Twister''~\cite{MaNMT} is the collective name of 
a family of PRNGs (pseudorandom number generators) based
on $\F_2$-linear maps.\footnote{Note that the $\F_2$ prefix will be often implied in
this paper.} This means that the state of the generator is a vector of bits of size $n$
interpreted as an $n$-dimensional vector on $\F_2$, the field with two elements, and the
next-state function of the generator is an $\F_2$-linear map. Since sum in
$\F_2$ is just xor, it is easy to implement such maps so that they can be
computed quickly. Several linear PRNGs indeed exist, such as
WELL~\cite{PLMILPGBLRM2} and \texttt{xorshift}~\cite{MarXR}.

The original paper about the Mersenne Twister was published by
Makoto Matsumoto and Takuji Nishimura in 1997~\cite{MaNMT}. At that time,
the PRNG had several interesting properties. In particular, it was easy
to build generators with a very large state space, and the largest version with
$19937$ bits of state became very popular. More importantly, many techniques
used in the Mersenne Twister influenced later development, and helped
$\F_2$-linear techniques to recover from the bad fame that followed the
``Ferrenberg affaire''.~\cite{FLWMCS}

It is difficult for non-specialists to
understand the intricacies of PRNGs, but period is easy to understand: the
fact that the sequence generated would not repeat before $2^{19937}-1$ $32$-bit
integers had been emitted was met with enthusiasm, and quickly the Mersenne Twister
was adopted as the standard generator in many environments. For example, the stock
PRNG of the \texttt{gcc} compiler and of Python, as well of the Maple mathematical
computing framework, is some version of the Mersenne Twister.

Nonetheless, since its very definition a number of problems plagued the Mersenne Twister.
First of all, it would have failed statistical tests designed to find bias in linear generator,
such as the Marsaglia's binary-rank test~\cite{MarTMSRNS} and the
linear-complexity test~\cite{CarALC,ErdETBK} (see Section~\ref{sec:lin}).
On the other hand in 1997 few
public, easy-to-use implementations of such tests were available, and finding bias at
the largest state sizes would have required an enormous amount of computing time.

It was also not clear to practitioners that $19937$ bits are ridiculously too much.
Even with as little as $256$ bits of state (and a period of $\approx 2^{256}$), the fraction of the output that can be
accessed by a computation is negligible, and even in a massive parallel
computation where processors start from random seeds the chances of overlap are
practically zero. In the last decades locality has increasingly become the main
factor in performance, and in retrospect using a generator wasting so many bits
of the processor cache did not make much sense; it does even less today.

In the following 20 years the Mersenne Twister had many reincarnations: a $64$-bit
version, a SSE2-based version~\cite{SaMSOFMT}, a version targeted to floating-point
numbers~\cite{MuMPSDPFTNUAT}, and so on. These versions provided more speed, and
improved somewhat some measures of quality, but they did not fix the problems in the
original version. In the meantime, research went on and several better alternatives
were discovered: PRNGs who would not fail statistical tests, had good theoretical
properties, and were in fact faster than the Mersenne Twister.

We are conservative with what we do not really understand. Once language
implementors decided to use the Mersenne Twister, the choice was doomed to be
cemented for a very long time. Also, the problems of the Mersenne Twister are
not immediately detectable in everyday applications, and most users really
interested in the quality of their PRNG will make an informed choice, rather
than relying on the stock PRNG of whichever programming environment they are
using. So there was no strong motivation to move towards a better PRNG.

This paper is mostly a survey on the known problems of the Mersenne Twister
family written for non-experts. We provide, besides reference to results in the
literature, some new, simple but insightful examples that should make even the
causal user understand some of the complex issues affecting the Mersenne Twister.
It is intended as a guide for casual users, and as well for language implementors,
to take an informed decision as to whether to use the Mersenne Twister or not.

\section{Failures in statistical tests for linearity} 
\label{sec:lin}

Since its inception, the Mersenne Twister was bound to fail two statistical tests that are 
typically failed by $\F_2$-linear generators: Marsaglia's binary-rank test~\cite{MarTMSRNS} and the
linear-complexity test~\cite{CarALC,ErdETBK}. All other linear generators, such as WELL~\cite{PLMILPGBLRM2}
and \texttt{xorshift}~\cite{MarXR}, fail the same tests.

These two tests exploit two different (but related) statistical biases. In the first case,
since the next state of the generator is computed by an $\F_2$-linear map
(a square matrix) $M$ applied to a state vector, the resulting bits,
when used to fill a large enough matrix on $\F_2$, yield a matrix that is not
random, in the sense that its rank is too low. In the second case, since every
bit of a linear generator is a \emph{linear-feedback shift register}
(LFSR)~\cite{KleSC} defined by the characteristic polynomial of $M$ (the only
difference between the bits is that they emit outputs at different points in the
orbit of the LFSR) the sequence emitted by such bits can be represented by a
linear recurrence of low degree,\footnote{More precisely, any sequence with period length $P$
can be represented with a linear recurrence of degree $P$, but in the case
of linear generators the degree is $O(\log P)$.} and this fact can be detected using the
Berlekamp--Massey algorithm provided that one uses a large enough degree upper bound
(and consequently memory) for determining the linear recurrence.

The two tests are partially orthogonal: that is, it is possible to create PRNGs
that fail one test, but not the other, under certain conditions.
On the other hand, they are also partially related, as a small matrix gives a
low-degree LFSR, which will fail easily the linear-complexity test, and causes
the generator to fail the binary-rank test with small matrices. Moreover, some
generators might have bits defined by different linear recurrences, so some bits
might fail the linear-complexity (and maybe the binary-rank) test while tested
in isolation, but not when mixed within in the whole output.

Since the two tests depend on a size (the matrix size, or the upper bound on the
degree of the linear recurrence), one should in principle always specify
this size when claiming that one of the tests is passed or failed: for example,
if some bits have higher linear degree than others a larger matrix might be necessary
to detect their linearity by means of the binary-rank test.

At this point, a good question is: How come that we are using in so many applications a generator that fails
statistical tests? There are two main reasons.

The first reason is that these tests were conceived \emph{specifically} to find
bias in linear generators, using knowledge of their internal structure. In this sense
they are different from classical tests, such as the \emph{gap test}~\cite{KeSRRSN}, which
look for generic undesired statistical regularities (or irregularities).

There is a thin line after which a statistical test becomes an attack, in the
sense that it uses too much knowledge about the generator. A good extreme example
is that of a PRNG based on a AES in counter mode~\cite{SMDPRN}: we fix a 
secret, and AES gives us (say) a bijection on $128$-bit blocks that is difficult
to reverse. We then apply the bijection to $128$-bit integers $k$, $k+1$, $k+2$,\,\ldots, and so
on, obtaining a strong random generator.\footnote{Modulo
some external entropy regularly introduced into the state of the generator, this
is what many secure PRNGs (such as Fortuna~\cite{FeSPC}) do.} Finding bias
in such a generator would mean, essentially, finding a weakness in the AES
encryption scheme.

However, we can easily design a statistical test that this generator will fail:
we simply apply to $128$-bit blocks of output the AES \emph{decoding}
function, and then apply a randomness test to the resulting sequence. Since for our PRNG the
resulting sequence is $k$, $k+1$, $k+2$\,\ldots\ for some $k$, it will fail every
statistical test.

Does this mean that AES in counter mode fails statistical tests? Really, no, because we
used an enormous amount of knowledge about the generator to design the test: basically,
we are cheating---nobody can invert AES that easily. One might argue that the same is true
of the binary-rank or of the linear-complexity tests, but, once again, we are walking a thin line. 

The second reason is that experts in the field have repeatedly stated that
failures in these tests are not harmful. For example, Pierre L'Ecuyer and Fran\c cois Panneton in their
survey on $\F_2$-linear generators~\cite{EcPFLRNG} state
\begin{quote}
All $\F_2$-linear generators fail the tests that look for linear relationships in the sequences of bits they produce, 
namely, the matrix-rank test for huge binary matrices and the linear complexity tests. [\ldots]
But whenever the bit sequences are transformed nonlinearly by the application (e.g., 
to generate real-valued random numbers from non-uniform distributions), the linear relationships between the
bits usually disappear, and the linearity is then very unlikely to cause a problem.
\end{quote}
The authors describe then a number of ways in which the result can be altered so that linearity
disappears.

The inventors of the Mersenne Twister
seem to be even less interested in such failures: in the
conclusion of the paper presenting the SFMT (SIMD-oriented Fast Mersenne Twister)~\cite{SaMSOFMT},
one of the last incarnations of the Mersenne Twister, 
Mutsuo Saito and Makoto Matsumoto discuss linear dependencies in the output in terms of \emph{equidistribution} (see Section~\ref{sec:ed}): 
\begin{quote}
Thus, it seems that $k(v)$ of SFMT19937 is sufficiently large, far beyond the
level of the observable bias. On the other hand, the speed of the generator is
observable. Thus, SFMT focuses more on the speed, for applications that require
fast generations. (Note: the referee pointed out that statistical tests based on
the rank of $\F_2$-matrix is sensitive to the linear relations [9], so the above
observation is not necessarily true.)
\end{quote}
Failures in statistical tests are not even mentioned in the paper, except for this note,
which was added at the request of a referee. In a subsequent paper on the dSFMT (double SIMD-oriented Fast Mersenne Twister)~\cite{MuMPSDPFTNUAT},
a version of the SFMT generating floating-point numbers with $52$ significant bits, they conclude:
\begin{quote}
They also passed TestU01 [11] consisting of $144$ different tests, except for
LinearComp (fail unconditionally) and MatrixRANK tests (fail if the size of
dSFMT is smaller then the matrix size). These tests measure the $\F_2$-linear dependency of the outputs, and reject $\F_2$-linear generators, such as MT, SFMT and
WELL.
\end{quote}
The failures are not discussed further: they are not deemed relevant.

Are these considerations correct? Certainly, if they are taken into their
context: linear generators \emph{should} not cause problems as long as their
output is used to generate uniform reals in the unit interval which are manipulated nonlinearly, as subsequent
operations will make the linear artifacts disappear. The scientists making these
claims have a background in simulation, where the main purpose of PRNGs is to
generate uniform deviates, which are then turned into some other deviate by
methods such as inversion or rejection~\cite{DevNURVG}, and, indeed, it is difficult
to imagine how linearity should influence the result in their case.

However, we are going to argue that, in fact, this is not the case in general.

\subsection{An instructive example}

Consider a mathematician (or some other kind of scientist) willing to understand
the structure of random binary square matrices. For instance, the mathematician might
be trying to understand the structure of the adjacency matrix $M$ of a directed Erd\"os--R\'enyi
graph~\cite{ERRG} with probability of a link $p=1/2$. Or the matrix describing a endorelation of
a set were $x$ relates to $y$ with probability $1/2$ (they are the same thing).

As it is customary, for example, in spectral graph theory, the mathematician might want to study some statistical property of 
the adjacency matrix, such as some property of its characteristic polynomial: she
would thus use a PRNG generate a few matrices, compute their characteristic polynomials and
have a look at the respective coefficients. As a first raw test, and also as a sanity check,
she might want to count the number of odd coefficients, which she should expect
to be about $n/2$, where $n$ is the side of the matrix. She would set up a PRNG,
use its bits to fill the matrices, duly compute their characteristic polynomials and check the resulting distribution.

We consider the following $64$-bit generators for this example:
\begin{itemize}
  \item AES in counter mode with $128$ bits of state, which we take as a reference, as it is a PRNG of cryptographic strength.
  \item \xorshiftp[128]~\cite{VigFSMXG}, a small generator based on an $\F_2$-linear map with $128$ bits of state,
  followed by a \emph{weak scrambler}~\cite{BlVSLPNG}
  which simply adds the two halves of the state. The resulting
  carry chains reduce the linearity of the bits as we
  progress from less significant to more significant bits: the
  lowest bit is generated by the same LFSR of the linear
  underlying generator, whereas higher bits see their linear
  degree increase enormously; moreover, there are no linear
  relationships between the bits~\cite{BlVSLPNG}.\footnote{This is the generator used by the Javascript engine of all major browsers.}
  \item \xoroshiropp[128],  a small generator based on an $\F_2$-linear map with $128$ bits of state,
  followed by a \emph{strong scrambler}~\cite{BlVSLPNG},
  which should delete all linearity.
  \item The SFMT (SIMD-oriented Fast Mersenne Twister)~\cite{SaMSOFMT}, one of the
  latest versions of the Mersenne Twister, with $607$ bits of state.
  \item WELL~\cite{PLMILPGBLRM2}, another $\F_2$-linear generator with excellent \emph{equidistribution}
  properties (in fact, it is \emph{maximally equidistributed}; see Section~\ref{sec:ed}), designed to improve the quality of the Mersenne Twister, with $512$ bits of state.
\end{itemize}

\begin{figure}
\centering
\includegraphics{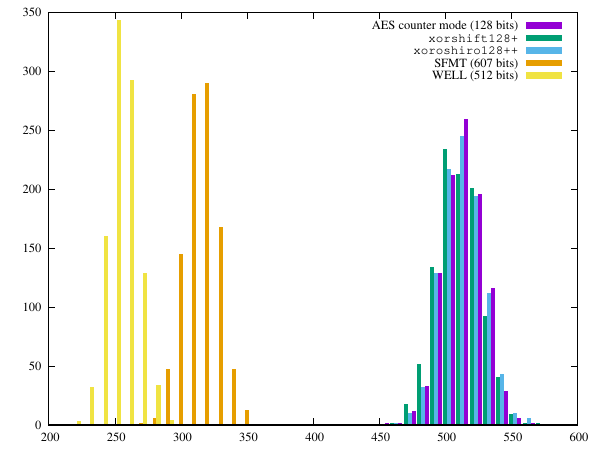}
\caption{\label{fig:poly0}The distribution of the number of odd coefficients in the
characteristic polynomial of $1024\times1024$ pseudorandom binary matrices generated by a few PRNGs using the output bits to fill
the matrix entries.}
\end{figure}
Figure~\ref{fig:poly0} shows the results of this simple test 
using a thousand samples and $n=1024$: as it is immediate to see, the results
of AES in counter mode and \xoroshiropp[128] are indistinguishable; their
averages are $512.9$ and $513.5$, respectively. The results
of the SMFT and of WELL are abysmal: the parity of the coefficients is completely wrong (the averages are $319.4$ and $258.1$, respectively).
Even \xorshiftp[128], despite emitting a few bits of low linear degree, is aligned with AES (average $512.1$).
If our friend mathematician would be using the SFMT (or WELL) for this purpose, she would be 
probably looking for some theoretical explanation of the strangely low number of odd terms.

Instead, she would be only witnessing the $\F_2$-linearity of the SFMT and WELL influencing
heavily the results, \emph{even if no $\F_2$-linear computation is involved}. The mathematically inclined
reader will probably have guessed what is happening: the matrices have low rank on $\F_2$, 
which means many eigenvalues are zeroes, which means that the characteristic polynomials
on $\F_2$ lack several coefficients, and the coefficients on $\F_2$ are just the projection
on $\F_2$ of the coefficients on $\Z$.

\begin{figure}
\centering
\includegraphics{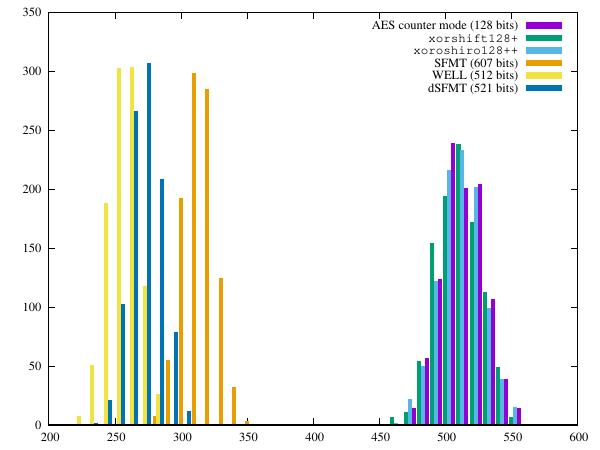}
\caption{\label{fig:poly1}The distribution of the number of odd coefficients in the
characteristic polynomial of $1024\times1024$ pseudorandom binary matrices generated by a few PRNGs using a $<1/2$ test on floating-point outputs.}
\end{figure}

One might think that the source of our problem is that we are using the output
of the generator directly, rather than generating a uniform number in the unit
interval and then manipulating it, as suggested in the references above. But this is not the case: we can perform
again the experiment, this time generating a random real number $p$ in the unit
interval and filling an entry with a $0$ if $p<1/2$, and with $1$ otherwise. Figure~\ref{fig:poly1}
was computed in this way and does not look much better, even if
we are using the PRNG to generate uniform random numbers in the unit interval.
In fact, we even added to the picture the dSFMT ($521$ bits), which generates floating-point numbers natively (see Section~\ref{sec:dSFMT}),
and its results are similar.

Readers
acquainted with the representation of floating-point numbers in the IEEE format
will immediately recognize that the test $p<1/2$ is simply using the most significant bit of the output of the generator:
in other words, we are using a single bit, rather than all bits, but
we are still filling the matrix with bits from the generator.

In fact, there's a small difference between the two figures: in the first figure,
\xorshiftp[128] has a slightly large variance than AES or \xoroshiropp[128]. The
variance however is the same in the second figure.
This happens because the most significant bit has the largest linear degree---in fact
it is essentially nonlinear. So this example shows also that some care must be taking
when evaluating linearity---in principle, not all bits are created equal. Generators
such as \xorshiftp[128] yield a nonlinear output when their upper $53$
bits are used to generate a uniform real number in the unit interval.

Still, one could think we are cheating, as clearly Panneton and L'Ecuyer speak of \emph{nonuniform} distributions,
and if we set or clear every element of the matrix using the probability threshold $p=1/2$ we are
dealing with a uniform distribution.

Thus, let us try to fill our matrices using the following nonuniform integer distribution:
\[
\Pr(2)=\frac14\qquad
\Pr(5)=\frac12\qquad
\Pr(6)=\frac14
\]
We will generate the distribution by a standard inversion, that is, we generate a real number in $[0\..1)$ and
emit $2$ if $p<1/4$, $5$ if $1/2\leq p < 3/4$, and $6$ if $p\geq 3/4$. Note that at this point we are
not even dealing with matrices with binary entries. 

\begin{figure}
\centering
\includegraphics{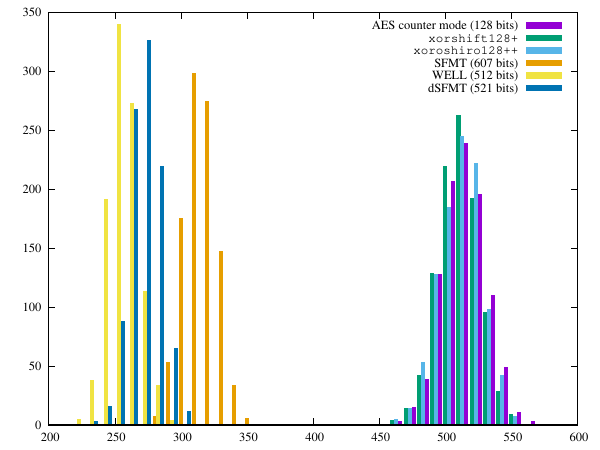}
\caption{\label{fig:poly2}The distribution of the number of odd coefficients in the
characteristic polynomial of $1024\times1024$ pseudorandom integer matrices generated by a few PRNGs using the distribution
$\Pr(2)=1/4$, 
$\Pr(5)=1/2$, $\Pr(6)=1/4$.}
\end{figure}

Alas, Figure~\ref{fig:poly2} does not show any improvement
on the previous cases. Once again, the reader acquainted with the IEEE representation of floating-point numbers will notice that we are
mapping real numbers with the most significant fractional binary digits $00$ or $11$ into even numbers, and the
others into odd numbers: thus, the \emph{parity} of the entries of the matrix
is given by the xor of the two most significant fractional binary digits,
and such a xor yields the LSFR associated with the generator: once again, the $\F_2$-linear artifacts of
the generators surface, even if we are performing integer-based computations using a non-uniform distribution. 

The lesson to be learned here is that generators should not fail statistical tests, at least
not easily.\footnote{\textit{Caveat}: ``Not easily'' needs to be specified as
every finite-state generator will fail every statistical test if enough time and space are allocated
for the purpose.} While a larger-state Mersenne Twister (or any kind of $\F_2$-linear 
generator with larger state) would need a correspondingly larger matrix
to show bias, the deviation from randomness is there, and its importance has been definitely downplayed,
as the bias can filter through computations that are not $\F_2$-linear.
Making a linear generator large and space-consuming just to avoid failing tests is not a good strategy.
 
\section{Failures in statistical tests for Hamming-weight dependencies}
\label{sec:hwd}

Linear generators have another known problem: if their next-state maps are represented by sparse matrices, states
with few ones or few zeros are usually mapped to states with few ones or few zeros. More generally, there
might be \emph{hidden dependencies} between the number of ones in the sequence of output emitted
by a linear PRNG. There are several tests which try to find such dependencies (e.g.,~\cite{LESBLCGMF,MaNNTWPNG}),
but the Mersenne Twister has no problem with them.

Recently, the author in collaboration with David Blackman has published a stronger test for
Hamming-weight dependencies based on a Walsh-like transform~\cite{BlVSLPNG}. We ran the test against
several generators until a petabyte of data has passed, or a $p$-value smaller than $10^{-20}$ was
computed: the sooner the low $p$-value appears, the stronger the statistical bias.
As one can see from Table~\ref{tab:testhwd}, all small-state members of the
Mersenne Twister family fail the test. For the standard Mersenne Twister we used 
Makoto Matsumoto's and Takuji Nishimura's library for dynamic creation of Mersenne Twisters~\cite{MaNDCPNG},
and we found a very wide range of sensitivity to the set: some dynamically generated
PRNGs requires a hundred times more data to fail the test than others.

As a reference, we report the same results for a $128$-bit
\texttt{xorshift}~\cite{MarXR} and \xoroshiro linear
generator~\cite{BlVSLPNG}.\footnote{Note that we do not suggest to use these
generators without scrambling~\cite{VigFSMXG,BlVSLPNG}.} The SFMT with $607$
bits of state fails using \emph{the same amount of data} of \xorshift[128], and
\emph{one order of magnitude less data} than \xoroshiro[128], both using a state
that is five times smaller: this should give a tangible idea of how sparse and
pathological the next-state function of the SFMT is. Linear generator with
significantly denser matrices such as WELL~\cite{PLMILPGBLRM2} need more data to
fail the test, but, as we remarked, they are an order of magnitude slower.
Strongly scrambled linear generators such as \xoshiropp[256]~\cite{BlVSLPNG} are
immune to these issues, they have sub-ns speed, and they have a reasonably sized
state.

Once again, larger-state Mersenne Twister PRNGs will not fail the test, as
catching dependencies in that case would require a large amount of memory. Nonetheless, as we already remarked, 
making a generator large and space-consuming just to avoid failing tests is not a good strategy.
 
\begin{table}
\renewcommand{\arraystretch}{1.3}
\begin{tabular}{lr}
Generator & \multicolumn{1}{c}{$p=10^{-20}$ @}\\
\hline
\xorshift[128]  & $8\times 10^8$ \\
\xoroshiro[128]  & $1\times 10^{10}$\\
Mersenne Twister ($521$ bits) & $4\times 10^{10}$--- \\
Mersenne Twister ($607$ bits) & $4\times 10^{8}$---$4\times10^{10}$\\
SFMT ($607$ bits) & $8\times 10^{8}$ \\
dSFMT ($521$ bits) & $7 \times 10^{12}$ \\
WELL512 ($512$ bits) & $3\times 10^{15}$ \\
\end{tabular}
\caption{\label{tab:testhwd}Detailed results of the test for Hamming-weight dependencies described
in~\cite{BlVSLPNG}. We report the number of bytes generating a
$p$-value smaller than $10^{-20}$.}
\end{table}

\section{Escaping from zeroland and decorrelation}

A known weakness of linear generators is the long time it takes, starting from a state in which there
are few ones (or few zeroes), to start generating bits which are equally likely zero or one. The reason,
once again, is that for the next-state map of a linear generator be computable quickly, its matrix
representation must be quite sparse. For example, the original Mersenne Twister ($19937$ bits) takes
hundreds of thousands of iterations, starting from a state in which exactly one bit is set, to get back to normality~\cite{PLMILPGBLRM2}.
This time is called sometimes ``escape from zeroland'': it was improved by the WELL
family of generators~\cite{PLMILPGBLRM2}, at the expense of a much slower generation (almost an order of magnitude).

While bad initial states are so rare that the probability of hitting such a bad sequence is 
negligible, for linear generators the escape time from zeroland is the same as the \emph{decorrelation} time:
given a state $S$ and a set $S'$ obtain by flipping a bit (or few bits), how much time it takes so that the 
two generated sequences are not correlated? Here by ``uncorrelated'' we mean ``empirically uncorrelated'',
exactly in the same way we consider a PRNG ``random'' when it is actually just ``empirically random'': to
test whether there is correlation, one just interleaves the sequences generate starting from $S$ and $S'$ and
uses a number of statistical tests to see whether the resulting sequence looks random.

A short decorrelation time is a good property, because it means that in the eventuality that two initial states
are similar, they will produce correlated sequences for a short time. From this viewpoint, linear generators with a large
state space, such as the Mersenne Twister with $19937$ bits of state, do not behave well---decorrelation takes
the same time as escaping from zero, as $V=S\oplus S'$ has just one bit set, so if $M$ is the next-state matrix by linearity $S M^k =S'M^k \oplus V M^k$:
as long as $VM^k$ contains mostly zeroes, $SM^k$ and $S'M^k$ will be correlated.

Note that while escaping from zeroland is a problem that is typical of linear generators, decorrelation is a
general problem for all fast generators with a large state space.
For example, we tested that Marsaglia's ``Mother-Of-All'' PRNG CMWC4096, which has a whopping period of $2^{131104}$, needs about
a dozen million iterations to decorrelate.

In general, large periods require a large state, and a large state cannot be perturbed too much
by a next-state function if you want speed. So the Mersenne Twister family is caught into a cross fire:
members with small state fail miserably linearity tests and tests on Hamming-weight dependencies. On
the other hand, members with a large state space might be able to pass some of those test, as
the large state space makes their linearity artifacts difficult to detect,
but they have very long decorrelation time, so similar states
will generate correlated sequences for a large number of iterations, and 
they occupy a large portion of the processor cache.

\section{Provable detectable deficiencies}

The extremely sparse nature of the next-state matrices of members of the Mersenne
Twister family is at the origin at the large ``escape from zeroland'' time we
already discussed, but it induces also other problems. Shin
Harase~\cite{HarFRMTPMG} has analyzed in detail the $\F_2$-linear relationship
induced by the structure of the Mersenne Twister with $19937$ bits of state, and
reports that subsequences with specific lags fail Marsaglia's \emph{birthday
spacing} test~\cite{MarCVRNG}. This is of course a statistical defect, as in
principle any subsequence should be as random as the original sequence, and we
could design a statistical test that extract a subsequence with the specified
lags; but it is definitely beyond the thin line we discussed in
Section~\ref{sec:lin}---the test would uses too much knowledge about the
generator.

\section{Equidistribution}
\label{sec:ed}

One of the motivations for the study and promotion of $\F_2$-linear PRNGs is that it is possible to prove 
that they satisfy \emph{equidistribution} properties. In fact, the very title of the first
Mersenne Twister paper contained the claim that it was $623$-dimensionally equidistributed~\cite{MaNMT}.

Equidistribution is a uniformity property of the output of the generator seen as a set of $d$-dimensional
points in the unit cube. One fixes a parameter $\ell$, the number of most significant bits considered (if unspecified,
the whole output),
and a dimension $d$. Then for each possible state of the generator one builds a $d$-dimensional
vector whose elements are blocks of upper $\ell$ bits in $d$ consecutive outputs, and checks that all such
vectors appear exactly the same number of times, when varying the state in all possible ways~\cite{EcPFLRNG}.

For example, when $\ell=1$ and $d=1$ we are just checking that in the output of the generator
half of the most significant bits are one and half are zeros. Instead, when $\ell=w$ and $d=2$, where $w$
is the width of the output of the generator, we are checking that every pair of $w$-bit values 
appears exactly the same number of times in the output of the generator. For this to happen, of course, $\ell d$ must 
not be greater than state size in bits. The \emph{equidistribution score} of a generator is the number of allowable pairs $(\ell,d)$
for which the count is wrong: zero is the best score, whereas the worst score depends on the state and output sizes.

Equidistribution is a property that can be measured for every PRNG, but since it requires examining all the output
it can be feasibly computed only on the smallest generators. However, in the case of $\F_2$-linear generators
the measurement requires just a bit of linear algebra, and can be performed easily even for large-state
generators.

This easiness has pushed equidistribution to the forefront of the design of $\F_2$-linear generators:
for example the WELL family contains generators that are \emph{maximally equidistributed}, that is,
the count is right for \emph{all} pairs $(\ell,d)$: their equidistribution score is zero. The improvement
of equidistribution scores is considered a quality of the SFMT, as we have read in Section~\ref{sec:lin}.
Another theoretical measure of quality is the number of terms of the characteristic
polynomial of the next-state matrix, which should be close to half the degree~\cite{ComHCRBS}.
The possibility of computing these theoretical measures is one of the main arguments
for linear PRNGs.

In fact, equidistribution scores are often compared with scores in the \emph{spectral test}, a mathematical
test on multipliers for linear congruential generators, which is the basis for finding multipliers of good quality~\cite{KnuACPII}.
The argument is that we need to find PRNGs with theoretical good properties, like good equidistribution scores
or spectral-test scores, and then test them.

There are several problems with this argument. The first, of general type, is logical: if these scores
provide good generators, why should we test them? They should be good, period. The truth is that these
scores provide some insight into the quality of the generators, but definitely cannot be used to prove
that the generator will pass a battery of tests.

There is one partial exception, however: as Knuth reports in detail~\cite{KnuACPII}, Harald Niederreiter has proved 
that linear congruential generators whose multipliers have a good spectral score \emph{will} pass a version
of the serial test~\cite{NieDPNGLCGII,NieDPNGLCGIII,NiePNOC,NieQMCPN}.
This is, in fact, an incredibly beautiful and strong result, which, in the end, motivates the application of the spectral test.

\emph{No such result is known for equidistribution}. It is a property that just ``looks nice'', but it has no 
connection with the results of a relevant statistical test. It was inspired by similar properties in
the context of \emph{quasi Monte--Carlo methods}~\cite{NieRNGQMCM}, where one, for example, generates sequences of points that fill uniformly
a space in order to approximate an integral. But the idea there is that one uses the entire sequence
of points, whereas when we use a PRNG we enumerate just a small fraction of its entire sequence of outputs. 
And, indeed, maximally equidistributed $\F_2$-linear generators
fail binary-rank and linear-complexity tests (and our test of Section~\ref{sec:lin}) like any other.

The author has also tried to find an empirical correlation between good equidistribution score in~\cite{VigEEMXGS},
by testing extensively the $2200$ possible parameterizations of Marsaglia's \texttt{xorshift} generators and
comparing the results of statistical testing with equidistribution score: once one restricts to reasonably good
generators, the correlation is extremely weak.

Another observation that suggests that equidistribution does not necessarily lead to particularly good generators
is that the equidistribution score is intrinsically unstable.
Indeed, if we take a maximally equidistributed sequence, no matter how long,
and we flip the most significant bit of a single element of the sequence, the new sequence
will have the \emph{worst possible} score (that is, the count will be wrong for all pairs).
This happens because, as it is easy to prove,
if $\ell'\geq \ell$, $d'\geq d$ and a generator is not equidistributed with parameters $\ell$ and $d$
then it is not equidistributed with parameters $\ell'$ and $d'$, either: by flipping one bit we make
the generator not equidistributed for $\ell=d=1$, which implies the lack of equidistribution for all 
possible other pairs.
For instance, by flipping the most significant bit of a single chosen value out
of the output of WELL1024a we can turn its equidistribution measure from zero to
$4143$---the worst possible value. But for any statistical or practical purpose the two sequences
are indistinguishable---we are modifying one bit out of $2^5(2^{1024}-1)$. These considerations
offer strong evidence that equidistribution will never be linked to the result of a feasible statistical test.

We must also mention a fact that is rarely reported in the literature: while we cannot prove that a generator
will \emph{pass} some statistical test because of a good equidistribution score, we can prove that it will \emph{fail}
one. The \emph{collision test}~\cite{KnuACPII} generates blocks of bits using the output of a generator and counts how many blocks have appeared
before, that is, \emph{collide} with a previous block. If
the blocks are made of $r$ bits, by the birthday paradox collisions start to happen after $O\bigl(\sqrt{2^r}\bigr)$ blocks have been examined, and their
approximate distribution is known~\cite{KnuACPII}, so we can check whether the empirical and the theoretical distribution
do match and obtain a $p$-value.

Consider now a generator with $w$-bit output and $kw$ bits of state that is $\ell=w$, $d=k$ equidistributed: that is,
every $k$-tuple of consecutive outputs appears exactly once (for example, WELL with $512$ bits of state has this property). If we perform
a collision test on such tuples (i.e., on blocks of $kw$ bits) using a short sequence of blocks of length $O\bigl(\sqrt{2^{kw}}\bigr)$ the generator will fail: collisions will
never appear, as they start to appear after $O\bigl(2^{kw}\bigr)$ blocks, instead of $O\bigl(\sqrt{2^{kw}}\bigr)$ blocks, have been examined,
because equidistribution
for $w$ bits in dimension $k$ implies that blocks will repeat only after the whole period of the generator
has been enumerated. While for most realistic generators it would be impossible to execute this test in practice, we can compute
analytically its negative result.

These considerations are (willingly) somewhat paradoxical, as a proper collision test 
using a sequence maximizing the power of the test~\cite{THCTCTP}, that is, using $\approx 1.25\cdot 2^{kw}$ blocks, would be failed with or without equidistribution:
there is just not enough state. Moreover, a non-equidistributed generator cannot produce all possible
outputs blocks of $kw$ bits, which can be arguably considered a non-random feature. And
while non-equidistributed generators \emph{might} be able to make collisions happen after $O\bigl(\sqrt{2^{kw}}\bigr)$
blocks, in practice the number of collision would have the wrong distribution.

All in all, equidistribution cannot be used, alone, as a reason to forget
statistical defects and failures in tests.
Richard Brent has made in the past similar considerations~\cite{BreMEHDS}, but
most of the observations above are new.

Equidistribution is the main reason why some researchers in the field do not modify the
output of a linear generator using some nonlinear function---it would make their
equidistribution-centered design difficult to motivate. But this gives us generators that
fail relevant statistical tests.

\section{Dynamic generation of parameters}

Parameters for the Mersenne Twister can be generated using a library published by 
Makoto Matsumoto and Takuji Nishimura~\cite{MaNDCPNG}. That is, given an integer
and a allowable state size in bits, the library provides parameters that can be used to instantiate
different Mersenne Twister. Having different parameters for a linear generator is
a standard situation: for example, Marsaglia's \texttt{xorshift} generators with
$64$ bits of state can be parameterized in $2200$ different ways~\cite{MarXR}. However,
both theoretical measures as equidistribution and statistical testing show that
such generators have a large variation in quality~\cite{VigEEMXGS}.

It is thus puzzling that specifically the Mersenne Twister would make it possible to
compute parameters that give generators always of high quality, as touted by the authors.
Indeed, this is not true: as we discussed in Section~\ref{sec:hwd}, different parameters
produced by the library yield generators which fail our test for Hamming-weight dependencies
with orders of magnitude of difference in the quantity of data analyzed. Even just
using the BigCrush suite from TestU01~\cite{LESTU01}, some of the generator fail only the usual linearity
tests, but other fail even the classical gap test~\cite{KeSRRSN}.

The conclusion is that the Mersenne Twister is not different from any other linear generator: while
we can compute parameters that satisfy some quality criteria and provide full period, we
do not have a complete understanding of the interaction between the parameters and the
statistical quality of the generator. Dynamic generation of parameters for the Mersenne
Twister is just dangerous.

\section{Direct generation of floating-point numbers}
\label{sec:dSFMT}

One of the latest addition to the Mersenne Twister family is the already mentioned dSFMT~\cite{MuMPSDPFTNUAT}, 
a generator that outputs directly floating-point numbers using a next-state \emph{affine}
function which leaves in the state of an almost exactly the representation of a
floating point number in IEEE format.

The dSFMT is very fast (sub-ns, if SSE2 instructions are available),
but it only generates $52$ significant bits. The IEEE format
can express $53$ bits of precision ($52$ in the significand, plus an implicit bit),
and thus the dSFMT can only generate half of the possible numbers in the unit
interval (assuming that by ``uniform generation of reals in the unit
intervals'' we mean generation of dyadics of the form $j/2^{53}$). Surprisingly,
this limitation is never discussed by the authors.

Finally, the dSFMT fails all tests for linear generators we discussed so far: in
view of the discussion of Section~\ref{sec:lin}, this case is even more
dangerous, because users not acquainted with linearity problems and with the
internal of the IEEE representation might think that the bias would not appear.
Nothing is farther from truth, as Figure~\ref{fig:poly1} and~\ref{fig:poly2} show.

Using a fast, reliable $64$-bit generator and turning its output into a floating-point
number is a much better and safer method: for example, \xoshirop[256] is a weakly
scrambled generators whose upper bits are nonlinear~\cite{BlVSLPNG}. By taking
the upper $53$ bits from its output and dividing them by $2^{53}$ one obtains 
a uniform floating-point number with none of the defects above. The generation
is $\approx8$\% slower than the dSFMT as the conversion to floating-point takes time, but
the advantages are overwhelming: no special instructions are needed, no tests
are failed, and one gets $53$ significant bits.

\section{Nonlinear Mersenne Twisters}

In sharp contrast with what we discussed up to this point, recently Mutsuo Saito and Makoto Matsumoto
have introduced a new PRNG, called the Tiny Mersenne Twister~\cite{SaMSFMT}, with the purpose
of providing a reliable generator with a small state ($127$ bits). Both a $32$-bit and
a $64$-bit exists.

The naming is at best confusing: the Tiny Mersenne Twister contains an addition, that is,
a nonlinear operation. It is thus not technically in the same family, and the name
``Mersenne Twister'' appear to have been used mainly for marketing purposes.
The nonlinear operation is necessary to obtain a generator with a small state that does not fail too many tests.

The Tiny Mersenne Twister passes the BigCrush test suite, but several of its bits are actually
of low linear degree or have linear dependencies: the lowest $32$ bits of the $64$-bit version, for example, 
or the whole output of the $32$-bit version, fail the binary-rank test,
and the lowest two bits fail the linear-complexity test (with the same parameters of BigCrush).
This is caused by the fact that a single nonlinear operation is not sufficient to delete all linear artifacts.

In fact, the situation is much worse: if we bit-reverse the output of the $32$-bit
version, the lower bits, which are statistically very weak, are analyzed by
BigCrush in detail, resulting in disastrous failures in numerous TestU01 tests such as
CollisionOver, SimpPoker, Run, AppearanceSpacings, LongestHeadRun, MatrixRank,
LinearComp and Run of bits~\cite{LESTU01}.
It is simply unfathomable why the authors would propose to use a PRNG so flawed.

Finally, the Tiny Mersenne Twister is very slow: the $64$-bit version needs almost 4\,ns
to emit a $64$-bit integer, whereas, for example, \xoroshiropp[128] needs less than a nanosecond, and
has none of its defects~\cite{BlVSLPNG}.

\section{Conclusion}

$\F_2$-linear generators fail statistical tests which can have an impact on actual
applications. They should be used with care, and only in context where there it
is certain that strongly nonlinear operations will be applied to their output,
so their defects are diluted or completely hidden. In particular, linear
generators should \emph{never} be used as general-purpose generators, unless their
output is suitably scrambled by combining it with other, nonlinear generators, or by 
applying nonlinear maps. The current, dangerous ubiquity of the Mersenne Twister as basic PRNG
in many environments is a historical artifact that we, as a community, should take care of.
Moreover, several touted advantages of the Mersenne Twister, such as reparameterization,
do not actually work properly when examined closely.

There are many available alternative to the Mersenne Twister which do not share
its defects. For example, the author has developed in collaboration with David
Blackman a number of generators formed by a linear engine and a
\emph{scrambler}---a bijection applied to the state of the linear engine that
tries to reduce or eliminate altogether linearity issues~\cite{BlVSLPNG}. Since
such generators do not need large state spaces to pass statistical tests, one
obtains general-purpose generators such as \xoshiropp[256] which have a
reasonable state space, sub-ns speed, and do not fail any known statistical
test. Of course the idea of scrambling is not new (e.g., it is suggested
in~\cite{LEPFRNG}), but we use a number of theoretical tools to prove properties
of our scramblers. Even \xoroshiropp[128], which we used in
Figure~\ref{fig:poly0}, \ref{fig:poly1} and~\ref{fig:poly2}, avoids linearity issues using just $128$ bits of state.

\bibliography{biblio}


\hyphenation{ Vi-gna Sa-ba-di-ni Kath-ryn Ker-n-i-ghan Krom-mes Lar-ra-bee
  Pat-rick Port-able Post-Script Pren-tice Rich-ard Richt-er Ro-bert Sha-mos
  Spring-er The-o-dore Uz-ga-lis }
\begin{thebibliography}{00}


\ifx \showCODEN    \undefined \def \showCODEN     #1{\unskip}     \fi
\ifx \showDOI      \undefined \def \showDOI       #1{{\tt DOI:}\penalty0{#1}\ }
  \fi
\ifx \showISBNx    \undefined \def \showISBNx     #1{\unskip}     \fi
\ifx \showISBNxiii \undefined \def \showISBNxiii  #1{\unskip}     \fi
\ifx \showISSN     \undefined \def \showISSN      #1{\unskip}     \fi
\ifx \showLCCN     \undefined \def \showLCCN      #1{\unskip}     \fi
\ifx \shownote     \undefined \def \shownote      #1{#1}          \fi
\ifx \showarticletitle \undefined \def \showarticletitle #1{#1}   \fi
\ifx \showURL      \undefined \def \showURL       #1{#1}          \fi

\bibitem[\protect\citeauthoryear{Blackman and Vigna}{Blackman and
  Vigna}{2019}]%
        {BlVSLPNG}
{David Blackman} {and} {Sebastiano Vigna}. 2019.
\newblock Scrambled Linear Pseudorandom Number Generators.
\newblock   (2019).
\newblock


\bibitem[\protect\citeauthoryear{Brent}{Brent}{2010}]%
        {BreMEHDS}
{Richard~P. Brent}. 2010.
\newblock \showarticletitle{The myth of equidistribution for high-dimensional
  simulation}.
\newblock {\em CoRR\/}  {abs/1005.1320} (2010).
\newblock


\bibitem[\protect\citeauthoryear{Carter}{Carter}{1989}]%
        {CarALC}
{G.~D. Carter}. 1989.
\newblock {\em Aspects of local linear complexity}.
\newblock Ph.D. Dissertation. University of London.
\newblock


\bibitem[\protect\citeauthoryear{Compagner}{Compagner}{1991}]%
        {ComHCRBS}
{Aaldert Compagner}. 1991.
\newblock \showarticletitle{The hierarchy of correlations in random binary
  sequences}.
\newblock {\em Journal of Statistical Physics\/} {63}, 5-6 (1991), 883--896.
\newblock


\bibitem[\protect\citeauthoryear{Devroye}{Devroye}{1986}]%
        {DevNURVG}
{Luc Devroye}. 1986.
\newblock {\em Non-uniform random variate generation}.
\newblock Springer Verlag.
\newblock


\bibitem[\protect\citeauthoryear{Erd\H{o}s and R\'{e}nyi}{Erd\H{o}s and
  R\'{e}nyi}{1959}]%
        {ERRG}
{P. Erd\H{o}s} {and} {A. R\'{e}nyi}. 1959.
\newblock \showarticletitle{On Random Graphs, I}.
\newblock {\em Publicationes Mathematicae (Debrecen)\/}  {6} (1959), 290--297.
\newblock


\bibitem[\protect\citeauthoryear{Erdmann}{Erdmann}{1992}]%
        {ErdETBK}
{E.~D. Erdmann}. 1992.
\newblock Empirical tests of binary keystreams.
\newblock   (1992).
\newblock


\bibitem[\protect\citeauthoryear{Ferguson and Schneier}{Ferguson and
  Schneier}{2003}]%
        {FeSPC}
{Niels Ferguson} {and} {Bruce Schneier}. 2003.
\newblock {\em Practical cryptography}.
\newblock Wiley.
\newblock


\bibitem[\protect\citeauthoryear{Ferrenberg, Landau, and Wong}{Ferrenberg
  et~al\mbox{.}}{1992}]%
        {FLWMCS}
{Alan~M. Ferrenberg}, {D.~P. Landau}, {and} {Y.~Joanna Wong}. 1992.
\newblock \showarticletitle{Monte {C}arlo simulations: Hidden errors from
  ``good'' random number generators}.
\newblock {\em Phys. Rev. Lett.\/}  {69} (Dec 1992), 3382--3384.
\newblock
Issue 23.


\bibitem[\protect\citeauthoryear{Harase}{Harase}{2014}]%
        {HarFRMTPMG}
{Shin Harase}. 2014.
\newblock \showarticletitle{On the $\mathbf{F}_2$-linear relations of
  {M}ersenne {T}wister pseudorandom number generators}.
\newblock {\em Mathematics and Computers in Simulation\/}  {100} (2014),
  103--113.
\newblock


\bibitem[\protect\citeauthoryear{Kendall and Smith}{Kendall and Smith}{1938}]%
        {KeSRRSN}
{Maurice~G. Kendall} {and} {Bernard~Babington Smith}. 1938.
\newblock \showarticletitle{Randomness and Random Sampling Numbers}.
\newblock {\em Journal of the Royal Statistical Society\/} {101}, 1 (1938),
  147--166.
\newblock


\bibitem[\protect\citeauthoryear{Klein}{Klein}{2013}]%
        {KleSC}
{Andreas Klein}. 2013.
\newblock {\em Stream Ciphers}.
\newblock Springer London, London.
\newblock


\bibitem[\protect\citeauthoryear{Knuth}{Knuth}{1998}]%
        {KnuACPII}
{Donald~E. Knuth}. 1998.
\newblock {\em The Art of Computer Programming, Volume 2: Seminumerical
  Algorithms\/} (third ed.).
\newblock Ad{\-d}i{\-s}on-Wes{\-l}ey, Reading, MA, USA.
\newblock


\bibitem[\protect\citeauthoryear{L'Ecuyer and Panneton}{L'Ecuyer and
  Panneton}{2005}]%
        {LEPFRNG}
{Pierre L'Ecuyer} {and} {Fran{\c{c}}ois Panneton}. 2005.
\newblock \showarticletitle{Fast random number generators based on linear
  recurrences modulo 2: overview and comparison}. In {\em Proceedings of the
  37th Winter Simulation Conference}. Winter Simulation Conference, 110--119.
\newblock


\bibitem[\protect\citeauthoryear{L'Ecuyer and Panneton}{L'Ecuyer and
  Panneton}{2009}]%
        {EcPFLRNG}
{Pierre L'Ecuyer} {and} {Fran{\c c}ois Panneton}. 2009.
\newblock \showarticletitle{$\mathbb F_2$-Linear Random Number Generators}.
\newblock In {\em Advancing the Frontiers of Simulation}, {Christos
  Alexopoulos}, {David Goldsman}, {and} {James~R. Wilson} (Eds.). International
  Series in Operations Research \& Management Science, Vol. 133. Springer US,
  169--193.
\newblock


\bibitem[\protect\citeauthoryear{L'Ecuyer and Simard}{L'Ecuyer and
  Simard}{1999}]%
        {LESBLCGMF}
{Pierre L'Ecuyer} {and} {Richard Simard}. 1999.
\newblock \showarticletitle{Beware of linear congruential generators with
  multipliers of the form $a=\pm2^q\pm2^r$}.
\newblock {\em ACM Trans. Math. Softw.\/} {25}, 3 (1999), 367--374.
\newblock


\bibitem[\protect\citeauthoryear{L'Ecuyer and Simard}{L'Ecuyer and
  Simard}{2007}]%
        {LESTU01}
{Pierre L'Ecuyer} {and} {Richard Simard}. 2007.
\newblock \showarticletitle{{TestU01}: {A} {C} library for empirical testing of
  random number generators}.
\newblock {\em ACM Trans. Math. Softw.\/} {33}, 4, Article 22 (2007).
\newblock


\bibitem[\protect\citeauthoryear{Marsaglia}{Marsaglia}{1985}]%
        {MarCVRNG}
{George Marsaglia}. 1985.
\newblock \showarticletitle{A current view of random number generators}. In
  {\em Computer Science and Statistics, Sixteenth Symposium on the Interface}.
  Elsevier, 3--10.
\newblock


\bibitem[\protect\citeauthoryear{Marsaglia}{Marsaglia}{2003}]%
        {MarXR}
{George Marsaglia}. 2003.
\newblock \showarticletitle{Xorshift {RNGs}}.
\newblock {\em Journal of Statistical Software\/} {8}, 14 (2003), 1--6.
\newblock


\bibitem[\protect\citeauthoryear{Marsaglia and Tsay}{Marsaglia and
  Tsay}{1985}]%
        {MarTMSRNS}
{George Marsaglia} {and} {Liang-Huei Tsay}. 1985.
\newblock \showarticletitle{Matrices and the structure of random number
  sequences}.
\newblock {\it Linear Algebra Appl.}  {67} (1985), 147--156.
\newblock


\bibitem[\protect\citeauthoryear{Matsumoto and Nishimura}{Matsumoto and
  Nishimura}{1998a}]%
        {MaNDCPNG}
{Makoto Matsumoto} {and} {Takuji Nishimura}. 1998a.
\newblock \showarticletitle{Dynamic creation of pseudorandom number
  generators}.
\newblock {\em Monte Carlo and Quasi-Monte Carlo Methods\/}  {2000} (1998),
  56--69.
\newblock


\bibitem[\protect\citeauthoryear{Matsumoto and Nishimura}{Matsumoto and
  Nishimura}{1998b}]%
        {MaNMT}
{Makoto Matsumoto} {and} {Takuji Nishimura}. 1998b.
\newblock \showarticletitle{Mersenne Twister: {A} 623-Dimensionally
  Equidistributed Uniform Pseudo-Random Number Generator}.
\newblock {\em ACM Trans. Model. Comput. Simul.\/} {8}, 1 (1998), 3--30.
\newblock


\bibitem[\protect\citeauthoryear{Matsumoto and Nishimura}{Matsumoto and
  Nishimura}{2002}]%
        {MaNNTWPNG}
{Makoto Matsumoto} {and} {Takuji Nishimura}. 2002.
\newblock {\em A Nonempirical Test on the Weight of Pseudorandom Number
  Generators}.
\newblock Springer Berlin Heidelberg, Berlin, Heidelberg, 381--395.
\newblock


\bibitem[\protect\citeauthoryear{Niedereiter}{Niedereiter}{1974}]%
        {NieDPNGLCGII}
{Harald Niedereiter}. 1974.
\newblock \showarticletitle{On the Distribution of Pseudo-Random Numbers
  Generated by the Linear Congruential Method {II}}.
\newblock {\em Math. Comp.\/}  {28} (1974), 1117--1132.
\newblock


\bibitem[\protect\citeauthoryear{Niedereiter}{Niedereiter}{1976}]%
        {NieDPNGLCGIII}
{Harald Niedereiter}. 1976.
\newblock \showarticletitle{On the Distribution of Pseudo-Random Numbers
  Generated by the Linear Congruential Method {III}}.
\newblock {\em Math. Comp.\/}  {30} (1976), 571--597.
\newblock


\bibitem[\protect\citeauthoryear{Niederreiter}{Niederreiter}{1977}]%
        {NiePNOC}
{Harald Niederreiter}. 1977.
\newblock \showarticletitle{Pseudo-random numbers and optimal coefficients}.
\newblock {\em Advances in Mathematics\/} {26}, 2 (1977), 99--181.
\newblock


\bibitem[\protect\citeauthoryear{Niederreiter}{Niederreiter}{1978}]%
        {NieQMCPN}
{Harald Niederreiter}. 1978.
\newblock \showarticletitle{Quasi-Monte Carlo methods and pseudo-random
  numbers}.
\newblock {\em Bull. Amer. Math. Soc.\/} {84}, 6 (1978), 957--1041.
\newblock


\bibitem[\protect\citeauthoryear{Niederreiter}{Niederreiter}{1992}]%
        {NieRNGQMCM}
{Harald Niederreiter}. 1992.
\newblock {\em Random number generation and quasi-{M}onte {C}arlo methods}.
  CBMS-NSF regional conference series in Appl.\ Math., Vol.~63.
\newblock SIAM.
\newblock


\bibitem[\protect\citeauthoryear{Panneton, L'Ecuyer, and Matsumoto}{Panneton
  et~al\mbox{.}}{2006}]%
        {PLMILPGBLRM2}
{Fran{\c c}ois Panneton}, {Pierre L'Ecuyer}, {and} {Makoto Matsumoto}. 2006.
\newblock \showarticletitle{Improved long-period generators based on linear
  recurrences modulo 2}.
\newblock {\em ACM Trans. Math. Softw.\/} {32}, 1 (2006), 1--16.
\newblock


\bibitem[\protect\citeauthoryear{Saito and Matsumoto}{Saito and
  Matsumoto}{2008a}]%
        {SaMSOFMT}
{Mutsuo Saito} {and} {Makoto Matsumoto}. 2008a.
\newblock \showarticletitle{SIMD-Oriented Fast Mersenne Twister: a 128-bit
  Pseudorandom Number Generator}.
\newblock In {\em Monte Carlo and Quasi-Monte Carlo Methods 2006}, {Alexander
  Keller}, {Stefan Heinrich}, {and} {Harald Niederreiter} (Eds.). Springer,
  607--622.
\newblock


\bibitem[\protect\citeauthoryear{Saito and Matsumoto}{Saito and
  Matsumoto}{2008b}]%
        {SaMSFMT}
{Mutsuo Saito} {and} {Makoto Matsumoto}. 2008b.
\newblock \showarticletitle{{SIMD}-oriented fast {M}ersenne {T}wister: a
  128-bit pseudorandom number generator}.
\newblock In {\em Monte Carlo and Quasi-Monte Carlo Methods 2006}. Springer,
  607--622.
\newblock


\bibitem[\protect\citeauthoryear{Saito and Matsumoto}{Saito and
  Matsumoto}{2009}]%
        {MuMPSDPFTNUAT}
{Mutsuo Saito} {and} {Makoto Matsumoto}. 2009.
\newblock \showarticletitle{A PRNG Specialized in Double Precision Floating
  Point Numbers Using an Affine Transition}.
\newblock In {\em Monte Carlo and Quasi-Monte Carlo Methods 2008}, {Pierre
  L'Ecuyer} {and} {Art~B. Owen} (Eds.). Springer Berlin Heidelberg, 589--602.
\newblock
\showDOI{%
\url{http://dx.doi.org/10.1007/978-3-642-04107-5_38}}


\bibitem[\protect\citeauthoryear{Salmon, Moraes, Dror, and Shaw}{Salmon
  et~al\mbox{.}}{2011}]%
        {SMDPRN}
{John~K. Salmon}, {Mark~A. Moraes}, {Ron~O. Dror}, {and} {David~E. Shaw}. 2011.
\newblock \showarticletitle{Parallel random numbers: as easy as $ 1, 2, 3 $}.
  In {\em {SC}'11: Proceedings of 2011 International Conference for High
  Performance Computing, Networking, Storage and Analysis, Seattle, {WA},
  November 12--18 2011}, {Scott Lathrop}, {Jim Costa}, {and} {William Kramer}
  (Eds.). ACM Press and IEEE Computer Society Press, 16:1--16:12.
\newblock


\bibitem[\protect\citeauthoryear{Tsang, Hui, Chow, Chong, and Tso}{Tsang
  et~al\mbox{.}}{2004}]%
        {THCTCTP}
{Wai~Wan Tsang}, {Lucas Chi~Kwong Hui}, {K.P. Chow}, {C.F. Chong}, {and} {C.W.
  Tso}. 2004.
\newblock \showarticletitle{Tuning the Collision Test for Power}. In {\em ACSC
  '04: Proceedings of the 27th Australasian Conference on Computer Science -
  Volume 26}. Australian Computer Society, Inc., 23--30.
\newblock


\bibitem[\protect\citeauthoryear{Vigna}{Vigna}{2016a}]%
        {VigEEMXGS}
{Sebastiano Vigna}. 2016a.
\newblock \showarticletitle{An experimental exploration of {M}arsaglia's
  \texttt{xorshift} generators, scrambled}.
\newblock {\em ACM Trans. Math. Software\/} {42}, 4 (2016).
\newblock
\newblock
\shownote{Article No.~30.}


\bibitem[\protect\citeauthoryear{Vigna}{Vigna}{2016b}]%
        {VigFSMXG}
{Sebastiano Vigna}. 2016b.
\newblock \showarticletitle{Further scramblings of {M}arsaglia's
  \texttt{xorshift} generators}.
\newblock {\it J. Comput. Appl. Math.}  {315} (2016), 175--181.
\newblock


\end{thebibliography}

\end{document}